\documentclass[aps,prl,showpacs,twocolumn,superscriptaddress, floatfix, tightenlines, amsmath, amssymb, longbibliography]{revtex4-1}

\usepackage{graphicx}
\usepackage{color}
\usepackage{dcolumn}
\usepackage{bm}
\usepackage{hyperref}
\usepackage[nolist,nohyperlinks]{acronym}
\usepackage{CJK}
\usepackage{braket}
\usepackage{nicefrac}
\usepackage{amsmath}
\definecolor{cblue}{RGB}{19,107,192}
\hypersetup{colorlinks=true,linkcolor=cblue,citecolor=cblue,urlcolor=cblue}

\newcommand{\GTS}{GaTa$_{4}$Se$_8$}
\newcommand{\GNS}{GaNb$_{4}$S$_8$}
\newcommand{\GNSe}{GaNb$_{4}$Se$_8$}
\newcommand{\GVS}{GaV$_{4}$S$_8$}
\newcommand{\GVSe}{GaV$_{4}$Se$_8$}

\newcommand{\cubsg}{F$\bar{4}$3m}

\newcommand{\pcubicsg}{P2$_{1}$3}

\newcommand{\orthorsg}{P2$_{1}$2$_{1}$2$_{1}$}

\begin{document}
\title{Jahn-Teller driven quadrupolar ordering and spin-orbital dimer formation in GaNb$_{4}$Se$_{8}$}

\author{Tsung-Han Yang}
\thanks{Current Affiliation: Neutron Scattering Division, Oak Ridge National Laboratory, Oak Ridge, TN 37831, USA}
\affiliation{Department of Physics, Brown University, Providence, RI 02912, USA}

\author{Tieyan Chang}
\affiliation{NSF’s ChemMatCARS Beamline, The University of Chicago, Advanced Photon Source, Argonne, Illinois 60439, United States}

\author{Yu-Sheng Chen}
\affiliation{NSF’s ChemMatCARS Beamline, The University of Chicago, Advanced Photon Source, Argonne, Illinois 60439, United States}

\author{K. W. Plumb}
\email[Corresponding author:]{kemp\_plumb@brown.edu}	
\affiliation{Department of Physics, Brown University, Providence, RI 02912, USA}

\date{\today}
\begin{abstract}
The lacunar spinel \GNSe{} is a tetrahedral cluster Mott insulator where spin-orbit coupling on molecular orbitals and Jahn-Teller energy scales are competitive. \GNSe{} undergoes a structural and anti-polar ordering transition at T$_{Q}=$~50~K that corresponds to a quadrupolar ordering of molecular orbitals on Nb$_4$ clusters. A second transition occurs at T$_{M}=$~29~K, where local distortions on the Nb$_4$ clusters rearrange. We present a single crystal x-ray diffraction investigation these phase transitions and solve the crystal structure in the intermediate $T_M\!<\!T\!<\!T_Q$ and low $T<T_{M}$ temperature phases. The intermediate phase is a primitive  cubic \pcubicsg{} structure with a staggered arrangement of Nb$_4$ cluster distortions. A symmetry mode analysis reveals that the transition at T$_Q$ is continuous and described by a single Jahn-Teller active amplitude mode. In the low temperature phase, the symmetry of Nb$_4$ clusters is further reduced and the unit cell doubles into an orthorhombic \orthorsg{} space group. Nb$_4$ clusters rearrange through this transition to form a staggered arrangement of intercluster dimers, suggesting a valence bond solid magnetic state.
\end{abstract}

\pacs{}
\maketitle
\section{Introduction}
The many interesting properties of strongly correlated magnetic materials arise from an interplay of charge, spin, orbital, and lattice degrees of freedom. In transition metal materials, it is often the case that spin and orbital effects appear in a hierarchy of well separated energy scales and can be considered to act independently of each other. For example, in transition metal materials that have an orbital degeneracy, the orbital degrees of freedom are  typically quenched out via a Jahn-Teller mechanism at a high temperature relative to the magnetic ordering temperatures, rendering a spin only magnetic model.   However, for 4d or 5d transition metal materials, relativistic spin-orbit coupling can significantly alter this energy hierarchy, giving rise to magnetic phases with spin-orbital entangled degrees of freedom. The presence of such spin-orbital degrees of freedom dramatically influences magnetic ground states and are the essential microscopic ingredient to stabilize, for example, Kitaev spin liquids, topological superconductors, or spin-orbital liquid phases depending on the particular electron filling~\cite{witczak-krempa:2014,  plumb:2014, romhanyi:2017, park:2020, tang:2023b}. In particular, 4d$^{1}$~or 5d$^{1}$~Mott insulators, with a single electron  occupying the $j=3/2$ spin-orbital state have attracted significant attention \cite{chen:2010, natori:2016, romhanyi:2017,iwahara:2018, natori:2017}  as materials where multipolar orders or spin-orbital liquids may be realized.
\begin{figure}[t!]
    \includegraphics{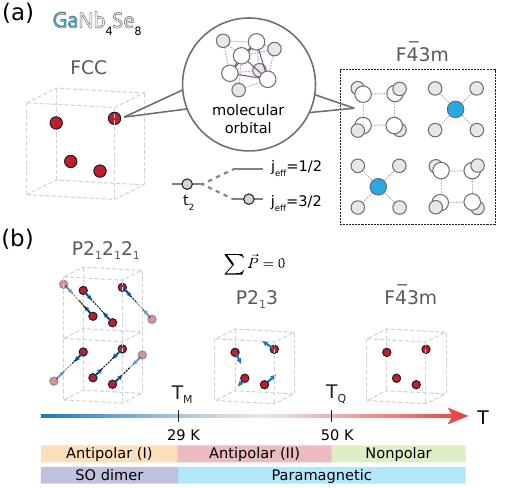}
\caption{(a) Crystal structure and molecular orbital level scheme \GNSe{}. The correlated units are molecular $j_{eff}=3/2$~orbitals on  Nb$_{4}$Se$_{4}$~ cubane units. Each Nb$_4$ cluster occupies an FCC lattice site. (b) Structural, magnetic, and polar phases transitions in \GNSe{} at different temperatures. The total polarization for two antipolar phases are zero while the internal arrangements differ.}
\label{fig:schematics}
\end{figure}

\begin{figure*}[ht!]
    \includegraphics{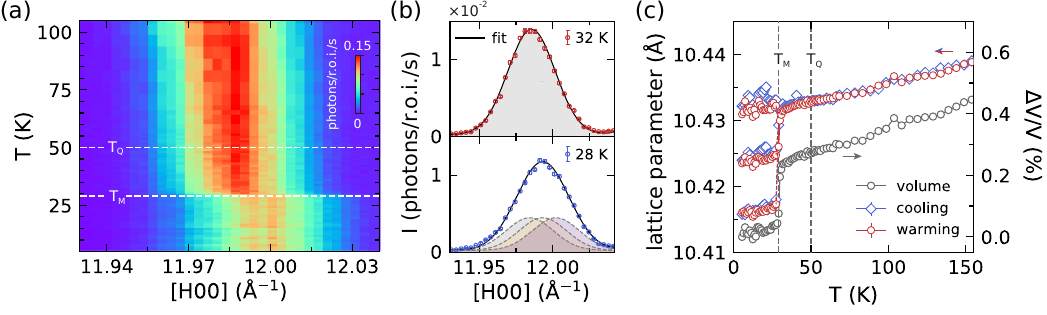}
\caption{(a) Temperature-dependent single crystal x-ray diffraction of the cubic [12,~0,~0] Bragg reflection in \GNSe{}. The [12,0,0] reflection is not sensitive to the transition at T$_{Q}$ but displays a clear shift and broadening below T$_{M}$ in the orthorhombic cell phase. (b) Gaussian fits of the [12, 0, 0] reflection above and below T$_{M}$. The peak width is fixed to instrumental resolution for the fit below T$_{M}$. (c) Temperature dependent lattice parameter(s) and the cell volume extracted from [12,~0,~0] reflection.}
\label{fig:LattParams}
\end{figure*}
Lacunar spinels with the chemical formula GaM$_{4}$X$_{8}$ (M=V, Nb, Ta; X=S, Se) are a family of cluster Mott insulators that contain a single unpaired electron occupying molecular $t_{2}$ orbitals on each M$_{4}$~cluster site [Fig.~\ref{fig:schematics} (a)] \cite{kim:2014a, kim:2020b}. The isostructural members of this family  that contain 3d, 4d, and 5d transition metals respectively provide an ideal opportunity to investigate the interplay of spin, orbit, and lattice degrees of freedom as atomic spin-orbit coupling and correlations are systematically varied. Indeed, clear trends in the relative separation of spin and orbital energy scales are observed across the series. In \GVS{} and \GVSe{} the orbital degrees of freedom are quenched through a high temperature Jahn-Teller (JT) distortion that is preceded by spin ordering at lower temperatures \cite{pocha:2000, wang:2015d}.  The JT distortion generated an electric dipole moment on each V$_4$ cluster that is oriented uniformly along the same direction for each cluster resulting in a ferro-polar ordering \cite{ruff:2017, butykai:2017, geirhos:2020b, geirhos:2021a, ghara:2021}.  On the other hand, in those compounds where spin-orbit coupling is the strongest, \GNS{} and \GTS{}, there is only a single magneto-structural transition leading to a dimerized, valence bond solid, ground state~\cite{arnold:2014b, ishikawa:2020}. The single transition, simultaneously involving spin and orbital degrees of freedom implies a vanishing separation between spin and orbital energy scales. Indeed, in  \GTS{}, the correlated units are molecular spin-orbit entangled $j_{\rm{eff}}=3/2$~\cite{kim:2014a, jeong:2017, yang:2022a, petersen:2023} degrees of freedom.  In such a spin-orbital state, JT effects are suppressed, but substantial spin-orbital dynamics are reflected in the disorder-order nature of the spin-orbital transition and corresponding phonon anomaly above the transition~\cite{yang:2022a}.  The structural distortions in \GNS{} and \GTS{} also generate electric dipoles on the transition metal clusters, but in these materials the distortions arrange in a staggered or anti-ferro polar ordering \cite{jakob:2007, geirhos:2021a, yang:2022a, winkler:2022}.

\GNSe{} presents an interesting intermediate regime where spin-orbit coupling and JT energy scales are on more equal footing. In this material, there is a cubic-to-cubic structural phase transition at T$_{Q}$=50~K preceding a magneto structural distortion \cite{pocha:2005, ishikawa:2020, winkler:2022} and the formation of structural dimers at T$_M=29$~K, illustrated in Fig.~\ref{fig:schematics}. The cubic-to-cubic structural distortion is associated with an anomaly in the temperature dependent dielectric constant \cite{winkler:2022} and the intermediate state suggests the possibility for anti-ferro electric quadrupolar ordering of $j=3/2$ spin-orbital degrees of freedom \cite{ishikawa:2020}. An additional anomaly of the dielectric constant at T$_M$ suggests a second rearrangement of electric quadrupoles at this temperature \cite{winkler:2022}. However, the precise form of quadrupolar ordering and underlying mechanism  in \GNSe{} have not been resolved because the crystal structures and atomic positions associated with the phases for T$_M~<$~T~$<$~T$_{Q}$ and T~$<$~T$_M$ are not known. This difficulty arises because of weak diffraction intensities from low temperature superlattice structural peaks, hindering the refinement of powder diffraction measurements, and from complications from domains in the low temperature orthorhombic unit cell. Detailed knowledge of the crystal structure, and especially the transition metal atomic positions, is essential to resolve the nature of quadrupolar or orbital ordering in this material and elucidate the microscopic mechanism driving this phenomena.

In this work, we investigate the temperature-dependent crystal structure of \GNSe{} using synchrotron single crystal x-ray diffraction. We determine the intermediate T$_M$~$<$~T~$<$~ T$_{Q}$ and low temperature T~$<$~T$_M$ crystal structures of \GNSe{} and find that the intermediate cubic structure is consistent with previous reports~\cite{ishikawa:2020}. However, a full refinement of the structure enables a symmetry analysis of the structural transition at T$_{Q}$. This analysis reveals it to be continuous (second order) transition described by a single JT active distortion of Nb$_{4}$ clusters. At T$_{M}=$~29~K, \GNSe{} further undergoes a first order cubic to orthorhombic transition. The unit cell doubles through this transition as Nb$_4$ clusters rearrange to form structural dimers along cubic $\langle$011$\rangle$ directions, consistent with the development of a correlated singlet, or valence bond solid state, that is similar to \GTS{} \cite{yang:2022a}. Our work sheds light on the complex interplay between spin-orbit coupling, lattice distortions, and intercluster interactions to effect electric quadrupolar orderings in Lacunar spinels and provides essential crystal structure information to support further investigation of spin-orbital physics in molecular Mott insulators.

\section{Methods}
\subsection{Sample Synthesis}
Polycrystalline samples of \GNSe{} were synthesized by solid-state reaction. About 1.5~g of stoichiometric quantities of high-purity raw materials (Gallium: Alfa Aesar, 99.999\%; Niobium: Alfa Aesar, 99.99\%; Selenium: Alfa Aesar, 99.999\%) were loaded into I.D.~=~10~mm, 100~mm long quartz ampules under an argon glovebox. The amupules were evacuated, and sealed before heating to 1050~$^{\circ}$C with 80~$^{\circ}$C/hour ramping rate, held for 24 hours and then air-quenched to room temperature. The resulting powder was reground in an argon glovebox, resealed under vacuum, and heated following the same procedure. Three heating and regrinding repetitions were required to obtain single-phase polycrystalline samples. The composition was confirmed by powder x-ray diffraction (Bruker Cu K$\alpha$ radiation). Micron scale single crystals of \GNSe{} were prepared by chemical vapor transport from the phase pure powders. About 1.5~g of single-phase \GNSe{} powder and 75~mg iodine were sealed in I.D.~=10~mm, 210~mm long quartz ampules. The ampule was placed in a two-zone furnace with 950~$^{\circ}$C and 1000~$^{\circ}$C zone temperatures for 20 days. 

\subsection{X-ray crystallography}
Temperature dependent x-ray diffraction was carried out using a four-circle x-ray diffractometer, Ag K$\alpha$ radiation,  with an XSpectrum Lambda 60~K GaAs photon counting detector positioned 78~cm from the sample. The single crystal sample was mounted in reflection geometry on a Cu post on the coldfinger of a closed cycle cryostat for measurements between 3 and 300~K.

Synchrotron crystallography measurements were performed on the 15 ID-D beamline at Advanced Photon Source at Argonne National Laboratory. The sample was mounted on a quartz fiber with epoxy adhesive. We used a 30~keV x-ray beam and Dectris Pilatus3X 1M CdTe area detector with a sample to detector distance of 130~mm for 100~K and 120~mm for 40~K and 20~K measurements. Nitrogen and helium cryo-streams were used to control the sample temperature. For each temperature, diffraction images were collected by rotating the sample through 360 degrees with 0.3 degrees/step for 100~K measurements and 0.1 degrees/step for 40~K and 20~K measurements. The images were integrated into reciprocal space and indexed using \textsf{Bruker APEX4} software. Absorption corrections were applied using \textsf{SADABS} and \textsf{TWINABS} packages and the space group was checked using the \textsf{XPREP} package. Final crystal structure solution and refinements were performed on \textsf{Shelx} with \textsf{Olex2} user interface \cite{sheldrick:2008, dolomanov:2009}.

\begin{figure}[t!]
    \includegraphics{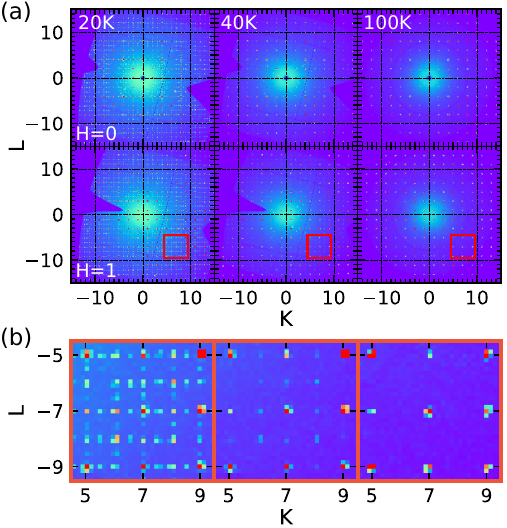}
\caption{(a) Reconstruction of (0KL) and (1KL) planes in reciprocal space from x-ray crystallography at 20, 40, and 100 K. (b) Zoomed areas in (1KL) plane, indicated by red boxes in (a), demonstrating the appearance of superlattice reflections.}
\label{fig:crystallography}
\end{figure}

\section{Results}
\subsection{Jahn-Teller driven antiferro quadrupolar order}
We first investigate the structural phase transition at T$_{Q}=$~50~K. Temperature-dependent diffraction measurements of a  cubic [12, 0, 0] Bragg peak are shown in Fig.~\ref{fig:LattParams}. The cubic peak exhibits noticeable broadening at T$_{M}=$~29~K, but no detectable change across T$_{Q}$ [Fig.~\ref{fig:LattParams}~(a)], consistent with a cubic-cubic phase transition. The temperature dependent lattice parameters and cell volume extracted from Gaussian fits of a longitudinal cut through the [12,~0,~0] peak are shown in Fig.~\ref{fig:LattParams} (c), demonstrating a continuous variation across T$_{Q}$, consistent with the second order nature of this transition.

Synchrotron x-ray crystallography at 40~K  reveals a primitive cubic unit cell and space group \pcubicsg{}. Fig.~\ref{fig:crystallography}~shows the appearance of reflections for $h+k=2n$; $h+l=2n$; $k+l=2n$ that violate reflection condition for a face-center-cubic (FCC) cell and indicate a primitive cubic unit cell at 40~K. Our refinement gives the best fit to the \pcubicsg{} space group with $R_{1}=$~0.0442 ($I \geq 2\sigma$). Nb atoms sit on $16e$ Wyckoff positions (WP) in the high temperature cubic phase that split into $4a$~and $12b$~WPs in the \pcubicsg{} structure resulting in an elongation of Nb$_{4}$ tetrahedra along the cubic $\langle \bar{1}11 \rangle$~directions as shown in Fig.~\ref{fig:xtl_structure} (a) and (b). Parameters of the refinement are listed in Table~\ref{tab:crystallography} and the refined atomic positions are listed in the supplementary materials~\cite{SM}. The space group and refined structure exhibiting a staggered arrangement of Nb$_4$ cluster distortions are consistent with anti-ferro quadrupolar ordering previously proposed in this regime \cite{winkler:2022, ishikawa:2020}.

\begin{figure}[t!]
    \includegraphics{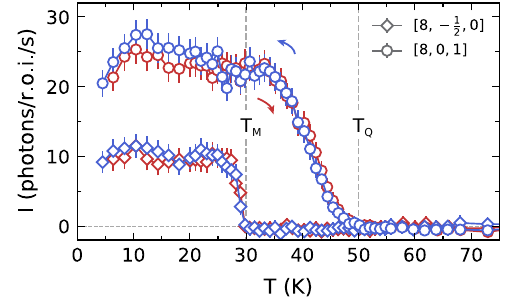}
\caption{Temperature-dependent peak intensities of and [8,~0,~1] Bragg peak that represents the order parameter for the \cubsg{} to \pcubicsg{} structural transition at T$_Q$=50~K, and [8,~$-\frac{1}{2}$,~0] reflection that shows a unit cell doubling through the first order transition at T$_{M}$=29~K.}
\label{fig:Transitions}
\end{figure}

We find that the transition at T$_{Q}$ is second order, consistent with the absence of any detectable volume change or hysteresis through the transition [Fig.~\ref{fig:LattParams} (c)], and a single irreducible representation order parameter for the distortion. The second order nature is apparent in Fig.~\ref{fig:Transitions},  that shows the continuous onset of the [8,~0,~1] reflection intensity at T$_Q$. This Bragg reflection is directly related to a single amplitude mode order parameter [Fig.~\ref{fig:xtl_structure} (b)] identified for this structural transition.

A symmetry mode analysis through the cubic-cubic \cubsg{}$\rightarrow$\pcubicsg{} distortion was carried out using the Bilbao crystallographic server~\cite{capillas:2003, orobengoa:2009, perez-mato:2010} to yield the primary mode amplitudes and isotrophy space groups listed in Table~\ref{tab:P213_dist}. Corresponding atomic distortions for each symmetry mode are listed in supplementary materials~\cite{SM}. The structural transition at T$_Q$ is described by two primary distortions: $\Gamma_{1}$, that belongs to the \cubsg{}, and a dominant JT active $X_{5}$ mode, that belongs to \pcubicsg{} and acts to distort Nb$_{4}$ clusters along $\langle \bar{1}11 \rangle$ directions as shown in  Fig.~\ref{fig:xtl_structure} (a) and (b). This second-order transition at T$_Q$ can be naturally described as a continuous increase of the $X_5$ mode amplitude order parameter [Fig.~\ref{fig:xtl_structure} (b)] that leads to the development of quadrupolar moments on Nb$_{4}$ clusters. Thus, a JT distortion of Nb$_4$ clusters underlies staggered quadrupolar ordering at T$_Q$ in \GNSe{}, similar to the proposed JT mechanism driving antipolar order in in \GNS{} \cite{geirhos:2021a}. However, it is interesting that in \GNSe{} the transition at $T_Q$ solely involves the ordering of structural (orbital) degrees-of-freedom, while magnetic and structural degrees of freedom undergo a second simultaneous transition at lower temperatures T$_M<$ T$_Q$. This separation of energy scales is distinct from \GNS{} and \GTS{} where there is only and single transition involving the structural and magnetic degrees of freedom.

\begin{table}[h]
\caption{\label{tab:DistAmpCC}Irreducible representation of distortion modes for \cubsg~to \pcubicsg~transition, amplitudes are normalized to the primitive unit cell 
of the high-symmetry \cubsg~structure.} \begin{ruledtabular}
\begin{tabular}{ c c c c c }
$\vec{K}$ & Irrep & Direction & Isotropy SG & Amp. (\AA) \\
(0,0,0) & $\Gamma_{1}$ & (a) & \cubsg & 0.0037 \\
(0,1,0) & $X_{5}$  & (a,a,a,a,a,a) & \pcubicsg & 0.0826 \\
\end{tabular}
\end{ruledtabular}
\label{tab:P213_dist}
\end{table}


\subsection{Structural dimers in the antipolar II ordered phase}
We now examine the evolution of the crystal structure through the first-order phase transition at T$_{M}=$~29~K, where a simultaneous structural transition and reduction in the temperature dependent magnetization occur. As shown in Fig.~\ref{fig:crystallography} and Fig.~\ref{fig:Transitions}, superlattice Bragg reflections, at [$\frac{1}{2}$,~0,~0] positions of the cubic cell, appear upon cooling below T$_M$. These indicate a unit cell doubling along one axis through a cubic to orthorhombic structural transition, similar to what has been observed in \GNS{} \cite{geirhos:2021a} and \GTS{} \cite{yang:2022a}.  The first order nature of this transition is identified though a discontinuous change in lattice parameters and cell volume [Fig.~\ref{fig:LattParams} (c)], a discontinuous onset of supperlattice reflection intensity [Fig.~\ref{fig:Transitions}], and sharp, divergent, heat capacity \cite{winkler:2022}.

\begin{figure}[t!]
    \includegraphics{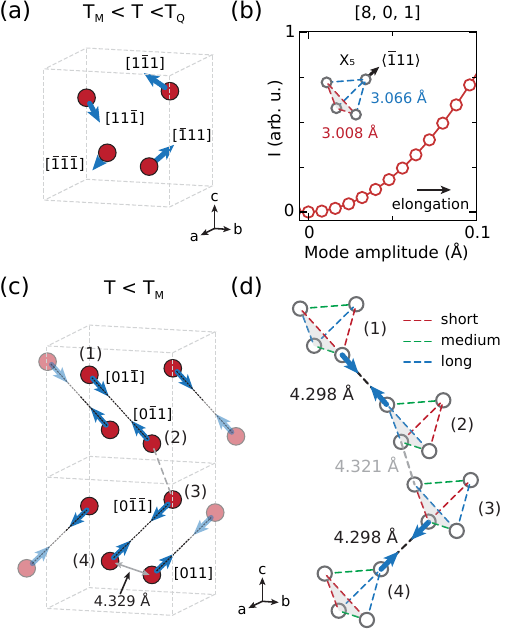}
\caption{ (a) Schematic representation of Nb$_4$ tetrahedron distortions in the T=40~K refined crystal structure. Red circles represent Nb$_{4}$~clusters, and blue arrows indicate the direction of overall distortion on each cluster in the cubic cell. (b) Simulated [8,0,1] peak intensity (using \textsf{Dans Diffraction} package \cite{Porter:2023}) and distortion on each Nb$_{4}$~cluster. The intensity of a [8,0,1] stuctural Bragg peak continuously increases with the $X_5$ distorton, as Nb$_4$ clusters elongate along $\langle \bar{1}$11$\rangle$ directions of the cubic \pcubicsg cell. (c) Schematic of Nb$_4$ cluster distortions in the  orthorhombic \orthorsg~cell at 20 K. Black dashed lines indicate the closest distance between neighboring clusters, showing a dimerized structure. (d) Schematic of the distortion of Nb$_{4}$~clusters. The location of each cluster within the orthorhombic cell are labeled with corresponding numbers in (c). Each cluster has two long bonds (blue), two short bonds (red) and two intermediate bonds (green).}
\label{fig:xtl_structure}
\end{figure}

In order to account for the formation of structural domains through this transition, we define four twinned cells for each doubled axis along the crystallographic {\bf a}, {\bf b}, or {\bf c} axis by rotating against the doubled axis through 0, 90, 180, and 270 degrees; giving 12 domains in total.  Integrated intensities of Bragg reflections were corrected based on symmetry operations using \textsf{TWINABS}~for these 12 domains. After accounting for the domain structure, our crystallographic refinement gives the best agreement with a unit cell doubled orthorhombic \orthorsg~space group with $R_{1}=$~0.0855 ($I \geq 2\sigma$). Refinement parameters are listed in Table~\ref{tab:crystallography}, and the refined structure is shown in Fig.~\ref{fig:xtl_structure} (c) and (d).

Fig.~\ref{fig:xtl_structure} (c) and (d) show the refined crystal structure and refined atomic positions are listed in the supplementary materials~\cite{SM}. At T$_M$, Nb$_{4}$ clusters distort so that each tetrahedron elongate along cubic $\langle$011$\rangle$~or equivalent directions [Fig.~\ref{fig:xtl_structure}~(c)] and has three inequivalent bond lengths as illustrated in Fig.~\ref{fig:xtl_structure}~(d). Distorted Nb$_4$ tetrahedra rearrange so that  each points towards a single neighboring cluster, forming structural dimers along $\langle$011$\rangle$~or $\langle$0$\bar{1}$1$\rangle$~directions as indicated in Fig.~\ref{fig:xtl_structure}~(c). The symmetry reduction on Nb$_4$ tetrahedra must have an associated reorientation of the polar axis, and their staggered arrangement indicates an antipolar ordering. Thus, the crystal structures reported here are consistent with the two successive antipolar orderings found at T$_Q$ and T$_M$ in \GNSe{} \cite{winkler:2022}, and reveal the nature of those orderings. The structural dimer motif and vanishing magnetization at T$_M$ \cite{ishikawa:2020, winkler:2022} also suggest magnetic singlets form across intercluster dimers as \GNSe{} enters a valence bond crystal like state below T$_M$, similar to what has been observed in \GTS{} \cite{yang:2022a}.

The unit cell doubled \orthorsg{} structure is not a subgroup of the intermediate \pcubicsg{} structure and the transition at T$_{M}$ cannot be described in terms of a JT active mode as for that at T$_{Q}$. The 1$\times$1$\times$1 orthorhombic \orthorsg{} is the exclusive sub group of the intermediate \pcubicsg~structure, while the superlattice reflections indicate the structure is a 1$\times$1$\times$2 cell [Fig.~\ref{fig:crystallography}]. Therefore, in order to better interrogate the mechanisms that underlie this transition, we compute structure relations against the high temperature \cubsg{} structure to obtain irreducible representations and distortion mode amplitudes listed in Table~\ref{tab:DistAmpCO}. The distortion again involves a JT active $X_{5}$ mode, but in this case there are other unit cell doubling distortions of comparable magnitude that modulate intercluster bonds and mix with the JT active modes. The necessity of multiple distortion modes to describe the transition at T$_M$ is consistent with the involvement of more than one order parameter including at least structural and magnetic degrees of freedom, and indicates that a JT mechanism is not the origin of the transition at T$_M$. An involvement of distortions that modulate intercluster bonds, and the formation of magnetic singlets indicates that the transition at T$_M$ may instead be driven by intercluster interactions, similar to \GTS{} \cite{yang:2022a}.

\begin{table}[h]
\caption{\label{tab:DistAmpCO}Irreducible representation of distortion modes for \cubsg~to \orthorsg~transition, amplitudes are normalized to the primitive unit cell 
of the high-symmetry \cubsg~structure.} \begin{ruledtabular}
\begin{tabular}{ c c c c c }
$\vec{K}$ & Irrep & Direction & Isotropy SG & Amp. (\AA) \\
(0,0,0) & $\Gamma_{1}$ & (a) & \cubsg & 0.0045 \\
(0,0,0) & $\Gamma_{3}$ & (a,b) & F222 & 0.0035 \\
(0,$\frac{1}{2}$,0) & $\Delta_{3}\Delta_{4}$ & (0,0,0,0,a,b,0,0,0,0,-b,a) & C222$_{1}$ & 0.0697 \\
(0,1,0) & $X_{3}$  & (0,a,0) & P$\bar{4}$m2 & 0.0294 \\
(0,1,0) & $X_{4}$  & (0,a,0) & P$\bar{4}$n2 & 0.0042 \\
(0,1,0) & $X_{5}$  & (0,0,a,b,0,0) & P2$_{1}$2$_{1}$2 & 0.1140 \\
($\frac{1}{2}$,1,0) & $W_{1}$  & (0,0,0,0,-a,a) & I$\bar{4}$2d & 0.0202 \\
($\frac{1}{2}$,1,0) & $W_{2}$  & (0,0,0,0,-a,a) & I$\bar{4}$2d & 0.0295 \\
\end{tabular}
\end{ruledtabular}
\end{table}


\section{Discussion}
It is instructive to compare the phase transitions in other isostructural and isoelectronic lacunar spinels. The particular sequence of phase transitions with decreasing temperature, and low-temperature crystal structures of GaM$_{4}$X$_{8}$~(M=V, Nb, Ta; X=S, Se) are set by an interplay between JT effects, that act to quench orbital momentum, and spin-orbit coupling on the molecular orbitals, that acts to maintain an orbital character. When the JT effect dominates (M=V) the orbital and magnetic sectors are distinct i.e. a high temperature JT structural transition quenches the orbital degrees of freedom and is preceded by spin ordering at a lower temperature~\cite{pocha:2000, yadav:2008, ruff:2015, ruff:2017}. In both \GVS{} and \GVSe{}, the JT distortion selects a ferro polar order with a charge dipole directed along  cubic $\langle$111$\rangle$ directions ~\cite{ruff:2015, ruff:2017}. In the opposite limit of dominant spin-orbit coupling, as is relevant for \GTS{}, the spin and orbital energy scales are not separate and the relevant degrees of freedom are spin-orbital entangled $j_{{\rm eff}}=3/2$ states \cite{jeong:2017, petersen:2023}. In this case, there is a single magneto-structural transition at T$^{*}$=50 K that cannot be described by a JT active mode \cite{yang:2022a} Below T$^{*}$, \GTS{} enters a valence bond-solid like ground state, and the transition at T$^{*}$ is driven by interactions between Ta$_4$ clusters \cite{yang:2022a} that sets up a staggered arrangement of  cluster distortions. 

\GNS{} and \GNSe{} sit at an intermediate spin-orbit coupling strength, where the interplay between SOC and the JTE becomes most apparent. Since electrons occupy molecular orbitals on cubane (M$_4$X$_4$; M=V,~Nb,~Ta, X=S,~Se) units in the lacunar spinels, the effective \emph{molecular} spin-orbit coupling is the relevant energy scale. Molecular orbitals are formed from a linear combination of atomic d-orbitals, so the effective spin-orbit coupling strength will increase with increasing transition metal atomic number, but decrease as the M-X orbital hybridization and effective size of molecular orbitals is increased. Thus, the effective spin-orbit coupling strength should be larger in \GNS{} compared with \GNSe{} because the broader spatial extent of Se orbitals will enhance M-X hybridization across cubane units. This expectation is borne out through the appearance of only a single transition involving the magnetic and structural degrees of freedom in \GNS{}, but a separation of energy scales and two transitions in \GNSe{}. However, unlike the Vanadium compounds, both transitions in \GNSe{}, at T$_Q$ and T$_M$, involve rearrangements of Nb$_4$ clusters and an associated change in the molecular orbital configuration. In other words, the JT transition at T$_Q$ does not act to quench orbital degrees of freedom but instead reduces degeneracy by imposing a staggered quadrupolar order involving spin-orbital moments. As previously pointed out \cite{ishikawa:2020}, such a transition is similar to the expected quadrupolar ordering predicted for $j=3/2$ double perovskites \cite{chen:2010}; however, we find that quadrupolar ordering in \GNSe{} is driven by a JT mechanism (ionic motion), rather than intersite orbital exchange.

The JT transition involving spin-orbital degrees of freedom favors an anti polar order in \GNSe{} \cite{winkler:2022}, with the staggered polar vector oriented along $\langle$111$\rangle$ directions rather than $\langle$111$\rangle$ ferro polar order  as in \GVS{} and \GVSe{} \cite{ruff:2015}. At lower temperatures, a non-JT mechanism results in a rearrangement of the Nb$_4$ clusters to orient staggered polar order along $\langle 110 \rangle$ directions and form structural dimers. This second transition is similar to the spin-orbital ordering transition at T$^{*}$ in \GTS{} in that it involves distortions that strongly modulate intercluster bonds, and results in an intercluster stuctural dimerization along $\langle 110 \rangle$ directions. Although the particular staggered pattern of dimers is different between these two compounds, it is likely that both the transition at T$^{*}$ in  \GTS{} and at T$_M$ in \GNSe{} are driven by intercluster spin-orbital interactions.

\section{Summary and Conclusions}
We have solved crystal structures of \GNSe{} in the intermediate T$_M < $ T $<$ T$_Q$ and low T $<$ T$_M$ temperature regimes and find that the transitions at T$_Q$ and T$_M$ involve distinct mechanisms. At T$_{Q}$, we find a continuous second-order Jahn-Teller, quadrupolar ordering, phase transition on Nb$_4$ tetrahedron. At T $<$ T$_{M}$, the unit cell doubles along one axis and forms a singlet dimer state, resembling the sibling compound \GTS{} \cite{yang:2022a}. Compared to previous studies on heat capacity, magnetic susceptibility, and dielectric measurements~\cite{ishikawa:2020, winkler:2022}, our results complete the thus far missing structural information for \GNSe{}. By identifying the structural mechanisms of staggered quadrupolar orderings, these results provide insights into material parameters controlling polar ordering in the orbitally degenerate lacunar spinels. In the future it would be interesting to directly interrogate the spin-and-orbital dynamics in these materials and explore how these energy scales relate to the high pressure superconducting phases in \GTS{} and \GNS{} \cite{pocha:2005, abd-elmeguid:2004, taphuoc:2013, park:2020}.

\begin{table*}
\caption{\label{tab:crystallography}Crystallographic refinement results at 20, 40 and 100~K.} \begin{ruledtabular}
\begin{tabular}{llll}
 Temperature (K) & 20 & 40 & 100\\
  Crystal system & orthorhombic & cubic & cubic\\
  Space group & P2$_{1}$2$_{1}$2$_{1}$ & P2$_{1}$3 & F$\bar{4}$3m \\
  a (\AA) & 10.4090(5) & 10.4071(6) & 10.40910(10) \\
  b (\AA) & 10.4086(5) & - & - \\
  c (\AA) & 20.8145(10) & - & - \\
  Data collection diffractometer & 15-ID-D, APS & 15-ID-D, APS & 15-ID-D, APS \\
  Absorption correction & Multiscan & Multiscan & Multiscan \\
  Reflections collected & 66929 & 17258 & 13368 \\
  Independent reflections & 16630 (R$_{int}$=0.0422) & 2489 (R$_{int}$=0.0273) & 464 (R$_{int}$=0.0340) \\
  F(000) & 3736.0 & 1401.0 & 1868.0 \\
  $\lambda$ (\AA, synchrotron) & 0.41328 & 0.41328 & 0.41328 \\
  2$\theta$ range for data collection (deg) & 2.544 to 50.64 & 3.218 to 49.048 & 3.94 to 47.77 \\
  Index ranges & $-18\le h \le 19$ & $-18\le h \le 18$ & $-17\le h \le 15$ \\
    & $-19\le k \le 19$ & $-13\le k \le 16$ & $-19\le k \le 17$ \\
    & $-38\le l \le 38$ & $-15\le l \le 18$ & $-17\le l \le 19$ \\
  Data, restraints, parameters & 16630/0/205 & 2489/0/41 & 464/0/12 \\
  Goodness of fit & 1.088 & 1.272 & 1.190 \\
  R$_{1}$, wR$_{2}$($I\ge2\sigma$) & 0.0855, 0.1777 &  0.0442, 0.0776 & 0.0088, 0.0200 \\
  R$_{1}$, wR$_{2}$(all) & 0.0910,  0.1794 & 0.0470, 0.0789 & 0.0088, 0.0200 \\
  Largest diff. peak/hole & 8.10/-7.15 & 4.92/-2.54 & 0.98/-0.96 \\
  
\end{tabular}
\end{ruledtabular}
\label{tab:refinement}
\end{table*}

\begin{acknowledgments}
Work at Brown University was supported by the U.S. Department of Energy, Office of Science, Office of Basic Energy Sciences, under Award Number DE-SC0021223.  This work is based on experiment performed at NSF’s ChemMatCARS Sector 15 that is supported by the Divisions of Chemistry (CHE) and Materials Research (DMR), National Science Foundation, under grant number NSF/CHE- 1834750. Use of the Advanced Photon Source, an Office of Science User Facility operated for the U.S. Department of Energy (DOE) Office of Science by Argonne National Laboratory, was supported by the U.S. DOE under Contract No. DE-AC02-06CH11357. 
\end{acknowledgments}

\bibliography{bib.bib}

\end{document}